# First results of the new endcap TOF commissioning at BESIII

**Zhi Wu**[a,b,*], **Shengsen Sun**[a,b,†], **Yuekun Heng**[a,b,‡], **Xiaozhuang Wang**[a,b,c], **Yongjie Sun**[b,c], **Cheng Li**[b,c], **Hongliang Dai**[a,b], **Ping Cao**[b,c], **Jie Zhang**[a,b], **Weijia Sun**[b,c], **Siyu Wang**[b,c], **Yun Wang**[b,c], **Xiaolu Ji**[a,b], **Jinzhou Zhao**[a,b], **Wenxuan Gong**[a,b], **Mei Ye**[a,b], **Xiaoyan Ma**[a,b], **Mingming Chen**[a,b], **Meihang Xu**[a,b], **Xiaolan Luo**[a,b], **Rongxing Yang**[a,b,c], **Qi An**[b,c], **Xiaoshan Jiang**[a,b], **Zhen-an Liu**[a,b], **Shubin Liu**[b,c], **Kejun Zhu**[a,b]

[a] *Institute of High Energy Physics(IHEP), Chinese Academy of Sciences,
Beijing 100049, China*

[b] *State Key Laboratory of Particle Detection and Electronics,
USTC-IHEP, China*

[c] *Department of Modern Physics, University of Science and Technology of China(USTC),
Hefei 230026, China*
    *E-mail*: wuz@ihep.ac.cn,sunss@ihep.ac.cn,hengyk@ihep.ac.cn

ABSTRACT: The upgrade of the current BESIII Endcap TOF (ETOF) is carried out with the Multi-gap Resistive Plate Chamber (MRPC) technology. The installation of the new ETOF has been finished in October 2015. The first results of the MRPCs commissioning at BESIII are reported in this paper.

---

[*] Corresponding author.
[†] Corresponding author.
[‡] Corresponding author.

# Contents

## 1. Introduction

The BESIII experiment [1] study $e^+e^-$ collisions in the $\tau$-charm energy region at the Beijing Electron and Positron Collider II (BEPCII) [2]. Particle identification is a fundamental tool in the data analysis since it help to disentangle specific processes inside the high multiplicity events. The original ETOF system [1], which mainly identify pions and kaons, consists of two disks of 48 pieces of plastic scintillating counters covering the polar angle region of $0.83<|\cos\theta|<0.96$ as shown in Fig.1. Each counter consists of fast scintillator blocks (BC204) readout by fine-mesh photomultiplier tubes (Hamamatsu R5924). The time resolution measured by ETOF detector is 138 ps for $\pi$'s [3], which cannot completely satisfy the higher precision requirement of physics. There are two main reasons for the worse ETOF time resolution: one is the multiple scattering of particles in the thick aluminium endcaps of the main drift chamber [4]; the other is the uncertainties of the particle positions hitting the scintillators.

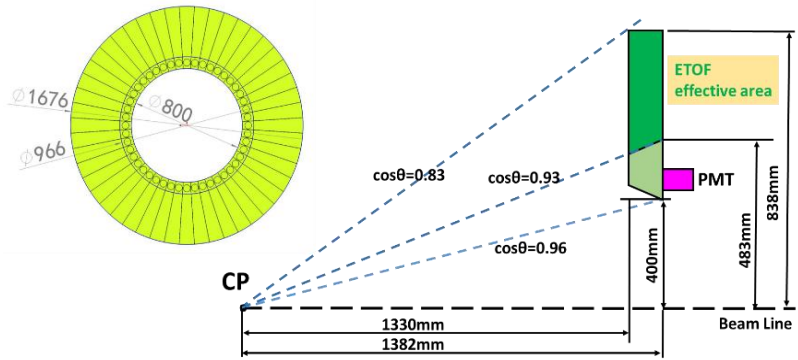

Fig.1 Schematic drawing of ETOF and location in BESIII.

The multi-gap resistive plate chamber (MRPC) [5] is a new type of gaseous detector with excellent time resolution, high detection efficiency, relatively low cost. It has been used successfully as TOF detectors in several experiments, such as ALICE at LHC [6][7], STAR at RHIC [8][9]. From 2011 to 2013, the R&D process has been carried out to develop the prototype MRPCs for the upgrade of BESIII ETOF. Test beam results have shown a time resolution better than 50 ps for 800 MeV pions which verify the physics design [10][11]. The segmentation of the MRPC can also be made sufficiently fine to suppress multi-hit events effectively at BESIII as discussed in Ref. [4]. In 2012 it was approved to replace the current BESIII ETOF with the MRPCs.

## 2. New ETOF detector

The new ETOF detector consists of 72 modules, 36 on each end as shown in Fig.2 (left). Adjacent modules are staggered to avoid dead regions. The effective areas of the MRPC rings have inner radius of 501 mm and outer radius 822 mm. Each MRPC is divided into 12 strips with readout from both ends. The readout granularity is increased by 12 fold compared to the original ETOF in which each plastic scintillator module is readout by only one PMT from the inner end. The lengths of readout strips in each MRPC module are from 9.1 cm to 14.1 cm and their width is 2.4 cm and separated by 3 mm, as shown in Fig.2 (right).

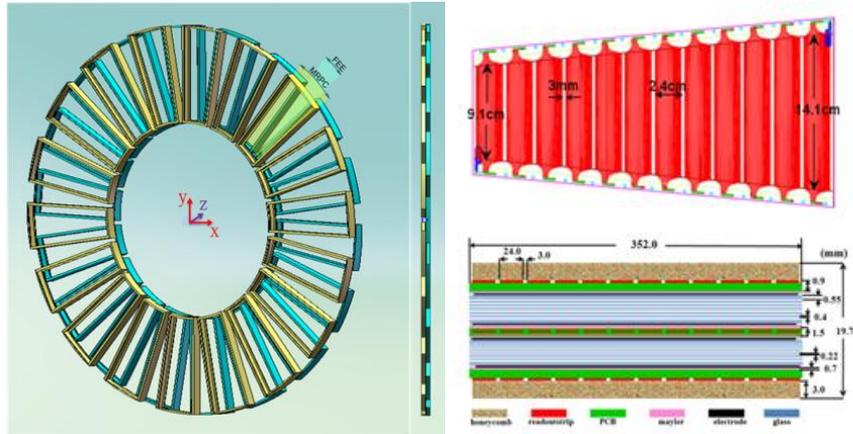

Fig.2 The schematic drawing of MRPC-based ETOF at BESIII (left); The layouts of printed circuit board and the cross-sectional view of a MRPC module along its length (right).

The gas system is made of 36 independent lines, 18 per end. Each line serves two modules serially. The gas mixture of Freon:$SF_6$:$C_4H_{10}$ (90:5:5) is used and controlled by the MKS 2179A&247D system. The high voltage system contains nine pairs of positive and negative channels. Each pair of channels are split to supply four modules and controlled by the BESIII slow control system.

## 3. Readout electronics and Online DAQ system

The readout electronics system of MRPCs consists of front end electronics (FEE) boards, Calibration-Threshold-Test-Power (CTTP) boards in NIM crate, Time- to- Digital (TDIG) converter modules, fast control modules and CLOCK modules in VME crate that works in pipeline mode with a clock frequency of about 40 MHz.

Each MRPC matches one FEE module with twenty four channels based on NINO [12], which features time-over-threshold (TOT) measurement. The timing accuracy is better than 15 ps RMS for each channel when the input charge is larger than 100 fC [13]. The FEE boards with connectors and cables are fixed on the surfaces of the aluminium boxes of the MRPC modules. Connectors (50 pin QSS-025-01-l-D-RA-MTI) and shielded differential cables (SQCD-025) are used to connect the FEE and the TDIG [14]. The CTTP board housed in a NIM crate provide power, threshold and test signal for the FEE. It also receives and transit the OR differential signals from the FEE to ETOF trigger subsystem.

The FEE signals are fed into TDIG modules, on each of which high performance TDCs [15] are utilized for precise time measurements. The L1 trigger is sent via fast control module to TDIGs. The CLOCK module receives the synchronized clock from the master, and fans out 18 clocks to feed modules inside the same VME crates.

The BESIII data acquisition system adopts the techniques of multi-level buffering, parallel processing, high-speed VME readout and network transmission. Also, the running status such as the event rate, noise level, the time spectrums and hitmap are monitored by a computer program to provide real-time information for the entire system.

Data quality monitoring system [16] is developed based on the offline software to monitor the accurate data quality in time. It can reconstructs part of the acquired data sampled randomly from online data flow. The performance on detector status can be available few minutes after physics run begins.

## 4. First results with collision data

The installation process lasts about two months, from August $1^{st}$ to October $7^{th}$ in 2015. The new ETOF commissioned together with BESIII since January 2016. Physics data-taking begins after the debugging of ETOF trigger sub-system. The preliminary results is described and discussed in the following sections. The Bhabha events are used to do the calibration.

### 4.1 Raw time distribution

As shown in Fig.3, a typical time distribution is obtained indicating that the MRPCs and the electronics work properly. The time window is set at 1600 ns, the same as that barrel TOF did. The hits in the region before the leading edge of the peak about 600 ns can reflect the noise level which will be given quantitatively in the following section.

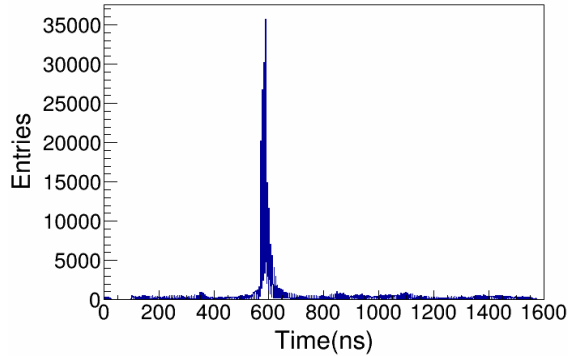

Fig.3 The time distribution.

### 4.2 Noise rate measurements

The single-channel rate is estimated by means of standalone files with a random clock trigger setting at a rate of 60 Hz during normal data taking. Since the TDIG has a multi-hit capability, at each event the number of hits in each readout channel is counted and divided by the time window of TDIG. At the nominal threshold value of 150 mV, the noise background distribution is shown in Fig.5, and an average rate of 9.0 Hz/cm2 is obtained.

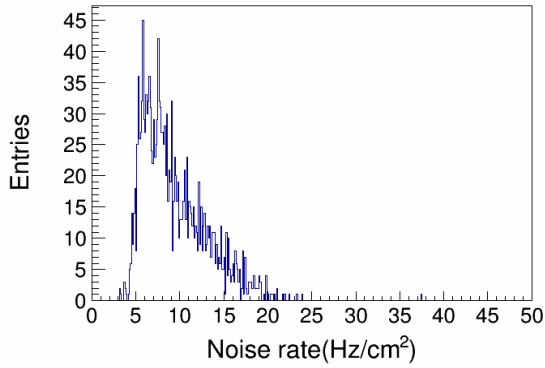

Fig.4 Noise rate distribution.

### 4.3 Event start time (EST) distribution

The trigger cycle is 24 ns, which equals the duration of 3 or 4 bunches. The TDC from ETOF is the time interval of the trigger start time to the arrival time of MRPC's signal. This time may differ from the interval between the collision time and the signal time in the detector. The interval of the trigger time and the collision time is called EST, calculated only by offline data analysis [17]. The BTOF or ETOF is the best choice to calculate the EST, since it near the tracker system and has the highest time measurement precision. The EST distribution calculated by ETOF is shown in Fig.5.

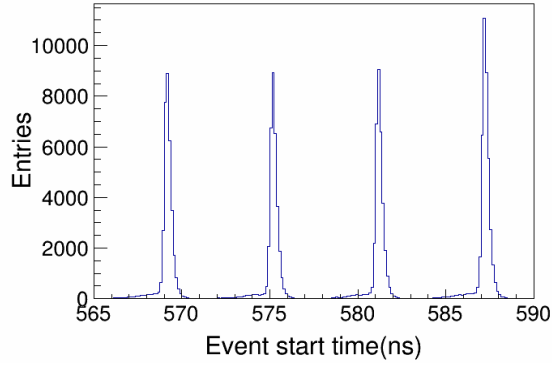

Fig.5 Event start time distribution.

### 4.4 Time resolution

The time calibration is carried out by comparing the measured time t_meas against the predicted time t_pred. The measured time t_meas is obtained by TDC subtracting EST, electronics offset, correction time caused by time-amplitude correlation effect. The predicted time t_pred is calculated by tracker system using the Kalman Filter method. The time resolution of 69ps for one-end (Fig.6 left) and 60ps for two-ends[1] (Fig.6 right) are determined by fitting the t_meas - t_pred distribution with Gauss function. Fig.7 shows the time resolutions for one-end and two-ends of MRPCs versus strip.

---

[1] The resolution of one-end stands for the result from the left or right channel in one strip; The resolution of two-ends stands for the combined result from left and right channels in one strip.

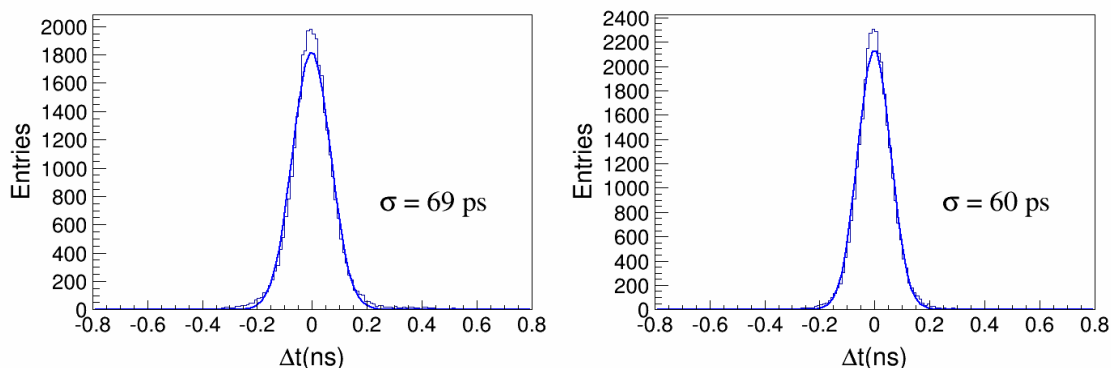
Fig.6 t_meas - t_pred distribution for one-end (left) and two-ends (right) of MRPC.

### 4.5 Detection efficiency

The detection efficiency is determined by the ratio of hits in one end or strip to the number of tracks passing through the strip. As shown in Fig.8, the average efficiency of two-ends is about 97.5%, which satisfies the design requirement of > 96%.

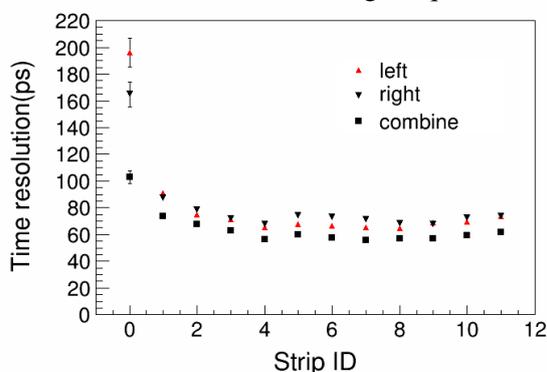
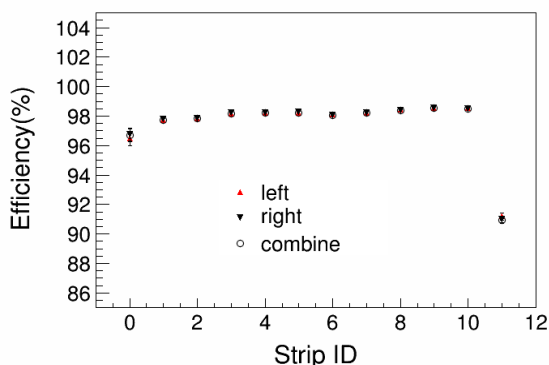

Fig.7 Time resolution versus strip.    Fig.8 Detection efficiency versus strip.

### 5. Conclusions

During the commissioning of the new ETOF, the status of all the hardware is stable. The noise rate has measured to be 9.0 Hz/cm$^2$, which reflects low noise level. After the calibration with Bhabha events, the very promising time resolutions of 69 and 60 ps for one-end and two-ends of MRPCs are obtained. The average detection efficiency of two-ends is about 97.5%. These preliminary results indicate the upgrade of BESIII ETOF is successful.

### Acknowledgments

This work is supported by the National Natural Science Foundation of China (No. 10979003) and Chinese Academy of Sciences (No. 1G201331231172010).